\documentclass[numberedappendix]{openjournal}
\usepackage{xcolor}
\usepackage{textgreek}
\usepackage[utf8]{inputenc}
\usepackage[english]{babel}
\usepackage{amsmath}

\usepackage{hyperref}
\hypersetup{
    unicode, 
    colorlinks=true,
    linkcolor=linkcolor,
    citecolor=linkcolor,
    filecolor=linkcolor,
    urlcolor=linkcolor,
}
\usepackage{color,colortbl}
\definecolor{linkcolor}{rgb}{0.0,0.3,0.5}
\usepackage{tensind}
\tensordelimiter{?}
\DeclareGraphicsExtensions{.bmp,.png,.jpg,.pdf}
\usepackage{verbatim}
\usepackage[normalem]{ulem}
\usepackage{orcidlink}
\usepackage{soul}

\graphicspath{ {./figs/} }

\begin{document}
\title{Accretion Disk Magnetic Braking}

\author{Kurt Liffman\orcidlink{0000-0002-4660-0162}}
\email{kliffman@swinburne.edu.au}
\affiliation{Centre for Astrophysics and Super Computing, Swinburne University of Technology, Hawthorn, VIC 3122, Australia}

\begin{abstract}
A protostellar disk is threaded by a static magnetic field that is perpendicular to the disk-surface. The magnetic field acts to brake the protostellar disk and cause the disk material to move towards the protostar. General analytic equations are derived for the accretion speed, and mass accretion rate.  Simplified analytic equations are also obtained for the disk energy dissipation, accretion timescale and the disk radial position plus disk surface density, as a function of time. In addition to providing physical insight, such equations might be useful as a check on computational models for protostar and protostellar disk formation.
\end{abstract}

% Write your keywords here

\maketitle

\section{Introduction}
\label{sec:int}

\begin{figure}
\centering
\includegraphics[width=\textwidth]{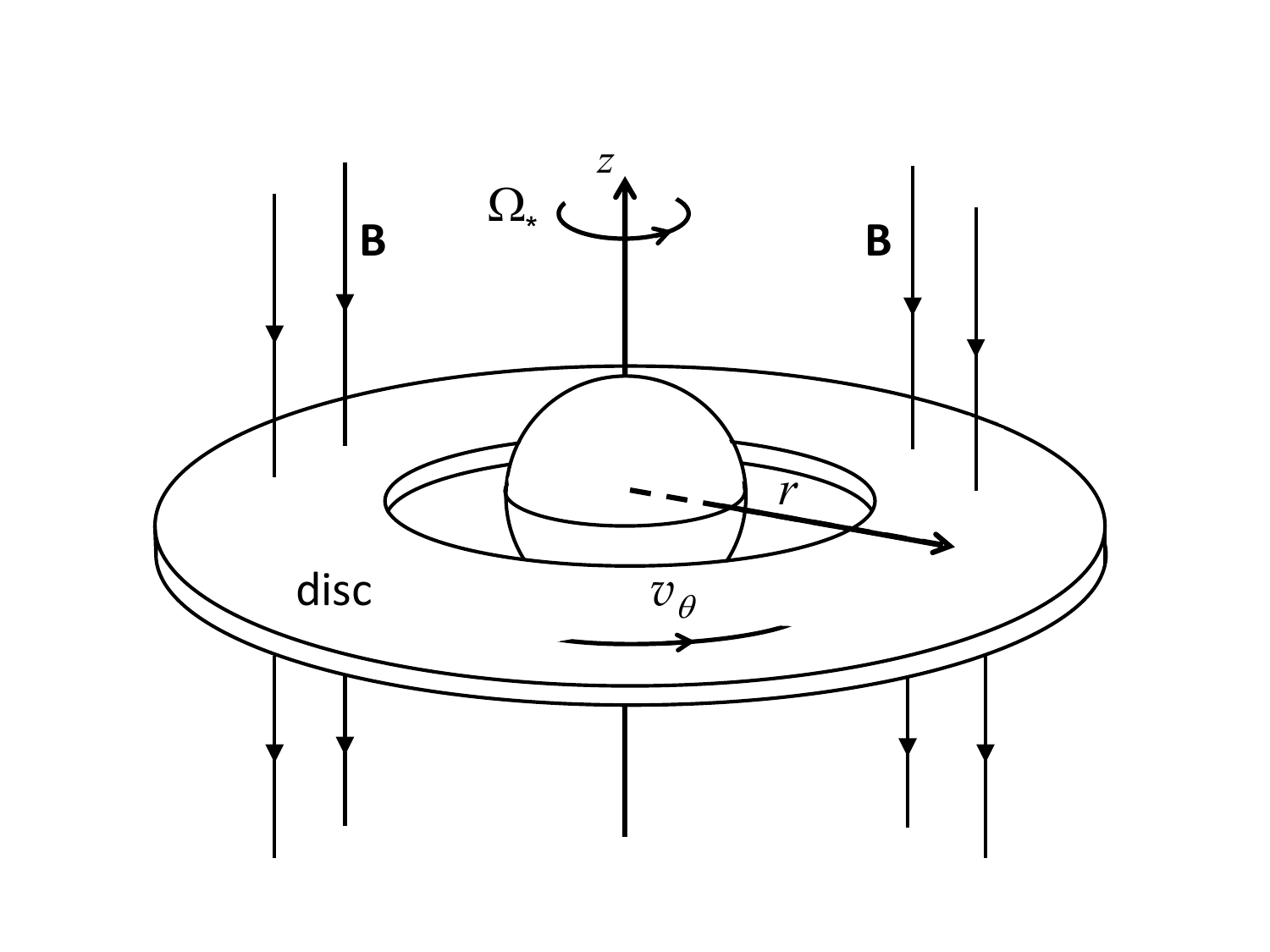}
\caption{A protostar rotates with an angular frequency $\Omega_*$, where the protostar is surrounded by a disk of material that has an angular velocity of $v_\theta$. The disk is threaded by a magnetic field $\mathbf{B}$ that is parallel to the axis of the system. Both $v_\theta$ and $\mathbf{B}$ may vary as a function of radial distance $r$ from the star. }
\label{fig:magnetic_disk}
\end{figure}

A typical scenario for the formation of a protostar is where a molecular cloud core collapses, and the cloud core is threaded by a magnetic field ({\it e.g.,} \citet{2012MNRAS.423L..45P}). The resulting protostellar system can be, initially, modelled as a protostar with surrounding accretion disk, where the disk is threaded by an axial magnetic field \citep{1994ApJ...433..746S} (Figure \ref{fig:magnetic_disk}).

For a partially conductive disk, the magnetic field has a braking action on the disk. This provides the disk with a radial infall speed and a rate of mass accretion towards the protostar. The braking action also causes the dissipation of disk energy. 

In this paper, we compute all these quantities with the assumption that the disk is inviscid ({\it i.e.}, any viscosity within the disk is negligible and can be ignored) and has a scalar electrical conductivity: $\sigma(r)$.

\section{disk Equations with Magnetic Braking}
\label{sec:diskeqn}

As discussed in \citet{2002apa..book.....F} , the equation for mass conservation in a disk with infalling material is:
\begin{equation}
   r \frac{\partial \Sigma}{\partial t} = -\frac{\partial \left(r v_{\rm r} \Sigma \right)}{\partial r} \ ,
    \label{eq:mass_con}
\end{equation}

with $r$ the radial cylindrical coordinate, $t$ the time, $\Sigma(r,t)$ the surface density of the disk, $v_{\rm r}$ the radial velocity of the flow of disk material.

Assuming the disk is inviscid, the angular momentum equation is tentatively:
\begin{equation}
   r \frac{\partial \left( \Sigma r^2 \Omega \right) }{\partial t} = -\frac{\partial \left( v_{\rm r}  \Sigma r^3 \Omega \right)}{\partial r} - 2h\sigma r^3 \Omega B^2 
   \ ,
    \label{eq:ang_con}
\end{equation}
where $\Omega(r,t)$ is the angular frequency of the disk, $h(r)$ is the disk scale height, $\sigma(r)$ the electrical conductivity of the disk and $B(r)$ the magnetic flux density. 

The first two terms of equation (\ref{eq:ang_con}) are the structural terms for the angular momentum conservation equation for a thin disk ({\it ibid.}). The third term provides an estimate for the loss of angular momentum due to magnetic braking (Appendix \ref{sec:magnetic_braking}). Angular momentum for the entire system might be conserved via twisting of the magnetic field and/or a disk wind. Such effects will not be considered here.

\subsection{Infall Speed}
\label{sec:vr}

As derived in Appendix \ref{sec:disk_infall}, the disk radial speed, $v_{\rm r}$, from equations (\ref{eq:mass_con}) and (\ref{eq:ang_con}), is:
\begin{equation}
    v_{\rm r}  = -\frac{2 h\sigma r^2 B^2 \Omega}{\Sigma \frac{\partial \left(  r^2 \Omega \right)}{\partial r}}
\label{eq:rad_drift1}
\end{equation}
For Keplerian angular disk speed, (i.e., $\Omega \approx \sqrt{{\rm G}M_*/r^3}$, where $G$ is the gravitational constant and $M_*$ the mass of the central object) we have:

\begin{equation}
    \begin{aligned}
    &v_{\rm r} = -\frac{4 h\sigma r B^2}{\Sigma} \\
     &\approx 4.7 \ \text{mm s}^{-1} \left(\frac{h}{0.09 \ \text{au}}\right) \left(\frac{r}{1 \ \text{au}} \right) \left(\frac{\sigma}{10^{-6} \ \text{S/m}}\right) \\
     &\times\left(\frac{B}{10^{-7} \ \text{T}}\right)^2 \left(\frac{\Sigma}{1.7\times10^{4}/\text{kg m}^{-2}} \right)^{-1} .
    \end{aligned}
\label{eq:rad_drift2}
\end{equation}
The infall speed, $v_{\rm r}$, is proportional to $B^2$ and so is, potentially, strongly dependent on this quantity. For a constant magnetic flux density, $v_{\rm r}$ decreases with $r$ as material moves towards the protostar. However, in protostellar systems, the magnetic fields tend to cluster near the protostar in a classic hourglass shape \citep{2024Univ...10..218B}. So, as one approaches the protostar, the expected increase in $\sigma$ and $B$ will tend to counter the decreases in $h$ and $r$.

\subsection{Mass Flow Rate}
\label{sec:dotM_a}

The mass flow rate or accretion rate, $\dot{M}_a $, within the disk has the form:
\begin{equation}
      \begin{aligned}
    \dot{M}_a &= 2 \pi r \Sigma v_{\rm r} = 8 \pi r^2 h \sigma B^2 \\
    &\approx 1.2\times10^{-9} \ \text{M}_\odot/\text{yr}\left(\frac{r}{1 \ \text{au}} \right)^2 
    \left(\frac{h}{0.09 \ \text{au}}\right) \left(\frac{\sigma}{10^{-6} \ \text{S/m}}\right) \\
     & \times \left(\frac{B}{10^{-7} \ \text{T}}\right)^2 \ .
    \end{aligned}
    \label{eq:Mass_flow_rate}
\end{equation}
This is a relatively small mass accretion rate, but semi-reasonable increases in the values of $\sigma$ and $B$ can significantly increase the accretion rate. 

\subsection{Energy Loss/disk Luminosity}
\label{sec:Energy_loss}

Magnetic braking produces energy loss, which can be dissipated as radiation at the disk surface. The disk power dissipation, $P_D$, is governed by the general equation (Appendix \ref{sec:disk_Power_Loss}):
\begin{equation}
    \frac{\text{d}P_D}{\text{d}r} = - 4 \pi \sigma r^3 \Omega^2 B^2 h \ .
\label{eq:disk_differential_powerII}
\end{equation}

So, if we assume keplerian rotation and constant values for $\sigma, h$ and $B$, then the total rate of energy dissipated by one side of the disk, $P_{D/2}$, is 
\begin{equation}
      \begin{aligned}
    P_{D/2} &\approx - 2 \pi \bar{\sigma} \bar{B}^2 \bar{h}\  {\rm G} M_* (r_{max} - r_{min}) \\
    &\approx 4.4\times 10^{-3} \ {\rm L}_\odot \left(\frac{\bar{h}}{0.09 \ \text{au}}\right) \left(\frac{\bar{\sigma}}{10^{-6} \ \text{S/m}}\right) \\
     & \times \left(\frac{\bar{B}}{10^{-7} \ \text{T}}\right)^2 \left( \frac{M_*}{{\rm M}_\odot} \right) \left( \frac{r_{max} - r_{min}}{100 \ {\rm au}} \right)\ .
        \end{aligned}
\label{eq:disk_one_side_power}
\end{equation}
with $\bar{\sigma}, \bar{h}$ and $\bar{B}$ the disk averaged values of $\sigma, h$ and $B$ plus
 $r_{max}$ and $r_{min}$ are, respectively, the maximum and minimum disk radii. 
 
 The disk luminosity is relatively small, but, again, the luminosity can be increased significantly with a plausible increase in the magnetic field density.

\subsection{Surface Density as a Function of Time}
\label{sec:Surface_Density}

As derived in Appendix \ref{sec:analytic solution}, the evolution equation for the disk surface density is

\begin{equation}
   \frac{\partial \Sigma}{\partial t} = \frac{1}{r}\frac{\partial \left(4h \sigma r^2 B^2 \right)}{\partial r} \ .
    \label{eq:mass_con2a}
\end{equation}
This can be solved ({\it ibid.}), via a rough, first approximation, as a disk surface density that evolves linearly in time
\begin{equation}
    \begin{aligned}
    \Sigma(r,t) & \approx \dot{\Sigma}_a (t-t_0) + \Sigma_0(r_0,t_0) \\
    &\approx 0.034 \ {\rm kg \ m}^{-2}{\rm yr}^{-1} \left(\frac{\bar{h}}{0.09 \ \text{au}}\right) \left(\frac{\bar{\sigma}}{10^{-6} \ \text{S/m}}\right) \\
     & \times \left(\frac{\bar{B}}{10^{-7} \ \text{T}}\right)^2\left( \frac{t-t_0}{\rm yr}\right) +\left( \frac{\Sigma_0(au,0)}{1.7 \times 10^4 \ {\rm kg \ m}^{-2}}\right)
   \ ,
    \end{aligned}       
    \label{eq:disk_surface_density_evolutionII}
\end{equation}
where $\dot{\Sigma}_a \equiv 8 \bar{h} \bar{\sigma} \bar{B}^2  $ is the rate of accretion surface mass density, with $\Sigma_0(r_0,t_0) \equiv \Sigma_0 $,  the initial value of $\Sigma$ at the initial position and time, $r_0$ and $t_0$.

In this scenario, as the disk accretes towards the protostar, the disk surface density, $\Sigma$, at a radial, co-moving point in the disk, will double in value after a time $\tau$, where the accretion timescale $\tau$ is
\begin{equation}
    \begin{aligned}
     \tau  &= \frac{\Sigma_0}{\dot{\Sigma}_a} \\ 
     &\approx 5 \times10^5 {\rm yr}
     \left( \frac{\Sigma_0(au,0)}{1.7 \times 10^4 \ {\rm kg \ m}^{-2}}\right) 
     \left(\frac{\bar{h}}{0.09 \ \text{au}}\right)^{-1} \\ 
     & \times \left(\frac{\bar{\sigma}}{10^{-6} \ \text{S/m}}\right)^{-1}
     \left(\frac{\bar{B}}{10^{-7} \ \text{T}}\right)^{-2}  \ .
    \end{aligned}
\label{eq:drift timescale II}
\end{equation}  

The disk surface density evolves on a timescale that is comparable to or somewhat smaller than the observed disk lifetimes of around $10^6 \ {\rm to} \ 10^7$ years \citep{2022ApJ...939L..10P}.

\subsection{Accretion Timescale and disk Position as a Function of Time}
\label{sec:Position_and_timescale}

Using the infall speed equation
\begin{equation}
    \begin{aligned}
    v_{\rm r} = \frac{{\rm d} r}{{\rm d} t}= -\frac{4 h\sigma r B^2}{\Sigma} ,
    \end{aligned}
\end{equation}

and the assumption of constant values of $\sigma$, $h$, and $B$, one can deduce the co-moving radial position of a point in the disk as a function of time (Appendix \ref{sec:analytic radial solution}):
\begin{equation}
    \begin{aligned}
    r & \approx  \frac{r_0}{ \sqrt{1 + (t-t_0)/\tau}} \ .
    \end{aligned}
\label{eq:radial position vs time II}
\end{equation}

We can invert this equation to find the time taken for the co-moving point to travel over a particular distance. For example, if we start at an initial radial position, $r_0$ at a time $t_0$ then the time taken for that point to reach a smaller radial distance $r_1$ at a time $t_1$ is:

\begin{equation}
    t_1 - t_0  \approx \tau \left( \left(\frac{ r_0}{ r_1 }\right)^2 - 1 \right) \ .
\label{eq:time vs radial position}
\end{equation}

According to equation (\ref{eq:time vs radial position}), if $r_1 = r_0/10$ then $t_1 - t_0 \approx 99 \tau$. Given that the value of $\tau$ might be of order $10^5$ years, then magnetic drag accretion does not appear to be a very promising mechanism for large-scale disk accretion. Of course, we have ignored the radial dependence of $\sigma$, $h$, and $B$, which may conspire to give small values for the accretion timescale, $\tau$, near the protostar.

\section{Conclusions}

\label{sec:conclusions}

Due to the complexity of the governing equations, theoreticians must employ sophisticated computational tools to study the formation and evolution of protostellar systems. Despite this necessary complexity, there is some utility in analysing aspects of these systems via relatively simple models. In this spirit, we have provided a brief analysis of an electrically conducting, protostellar disk that is subject to magnetic braking via a static magnetic field, where the field is perpendicular to the disk-surface. This situation is thought to arise during the early stages of protostar formation.

The applied magnetic field removes angular momentum from the disk and allows disk material to fall towards the protostar. This allows the derivation of general analytic equations for the radial disk accretion speed, and the corresponding mass accretion rate.  For the chosen representative values of the electrical conductivity of the disk, scale height, and magnetic flux density, the accretion speed and mass accretion rate are not significant. However, both quantities are proportional to the square of the magnetic flux density. So, near the protostar the accretion speed and accretion rate may increase to astrophysically relevant values, due to a likely increase in the magnetic flux density. 

Simplified analytic solutions are also obtained from more general differential equations that describe the energy dissipation, accretion timescale, and the disk radial position plus disk surface density, as a function of time. These solutions suggest that whole disk accretion due to magnetic drag may not be astrophysically relevant. However, magnetic braking may be relevant for disk gap formation and the inner regions of the disk near the protostar. 

In addition to providing physical insight, such equations might also be useful as a check on computational models for protostar and protostellar disk formation.

\section*{Acknowledgments}

I thank the Swinburne University's Centre for Astronomy and Supercomputing for its support.

\bibliographystyle{apsrev4-1}

% You should give the same name for your .bbl as your main .tex
% since it is a requirement for posting on ArXiv.
\bibliography{oja_template}

\begin{appendix}

\section{ Magnetic Braking}
\label{sec:magnetic_braking}

From Figure (\ref{fig:magnetic_disk}), the disk current density, $\mathbf{j}_D$, generated by the interaction between the disk and the external magnetic field is

\begin{equation}
\begin{aligned}
    & \mathbf{j}_D \approx \sigma (\mathbf{v}_\theta \times \mathbf{B})\ = \sigma (r\Omega \ \hat{\theta} \times (-B \ \hat{\mathbf{k}})) \\
    & = -\sigma r\Omega B \ \hat{\mathbf{r}},
    \label{eq:disk_current}
\end{aligned}
\end{equation}
where $\hat{\theta}$, $\hat{\mathbf{k}}$, and $ \hat{\mathbf{r}}$ are, respectively, the unit vectors in the $\theta$, $z$ and $r$ directions.

The Lorentz force per unit volume, $\mathbf{F}_D$, is
\begin{equation}
    \mathbf{F}_D = \mathbf{j}_D \times \mathbf{B} = -\sigma r\Omega B^2 \ \hat{\theta}.
    \label{eq:Lorentz_Force}
\end{equation}

If we take an annular volume element, $\Delta V$, of the disk, where the volume element has a radius $r$, a width $\Delta r$ and a height $2h$. Then $\Delta V = 4\pi r h \Delta r$ and the torque $\Delta \tau$ on the volume element is

\begin{equation}
\begin{aligned}
    & \Delta \tau = \mathbf{r}\times\mathbf{F}_D \ \Delta V \\
    & = -\sigma r^2\Omega B^2 \ (\hat{\mathbf{r}} \times \hat{\theta}) \ 4\pi r h \Delta r \\
    & = - 4 \pi \sigma r^3 \Omega B^2 h \Delta r \ \hat{\mathbf{k}}
\label{eq:disk_Torque}
\end{aligned}
\end{equation}

The torque per unit surface area:
\begin{equation}
    \frac{\Delta \tau}{2\pi r \Delta r} 
     = - 2 h \sigma r^2 B^2 \Omega  \ \hat{\mathbf{k}}
\label{eq:disk_Torque_per_area}
\end{equation}

is the third term in angular momentum conservation equation (equation (\ref{eq:ang_con})) prior to multiplication by $r$.

\section{ disk Accretion Speed}
\label{sec:disk_infall}
From equation (\ref{eq:mass_con}), 
\begin{equation}
    \frac{\partial(v_{\rm r} \Sigma r^3 \Omega)}{\partial r} = -r \frac{\partial \Sigma}{\partial t}r^2 \Omega  + r v_{\rm r} \Sigma \frac{\partial (r^2 \Omega)}{\partial r} \ ,
\end{equation}
while
\begin{equation}
\begin{aligned}
\frac{\partial \left( \Sigma r^2 \Omega \right)}{\partial t} &= 
\frac{\partial \Sigma}{\partial t} r^2 \Omega \ .
\end{aligned}
\end{equation}

Combining these equations with equation (\ref{eq:ang_con})
\begin{equation}
\begin{aligned}
    r\frac{\partial \left( \Sigma r^2 \Omega \right)}{\partial t} &+ \frac{\partial \left(r v_{\rm r} \Sigma r^2 \Omega \right)}{\partial r}   \\
    &=  r v_{\rm r}\Sigma\frac{\partial \left(  r^2 \Omega \right)}{\partial r} \\
    &=  - 2 h\sigma r^3 B^2 \Omega \ ,
    \end{aligned}
\end{equation}

and so,

\begin{equation}
     v_{\rm r}  = -\frac{2 h\sigma r^2 B^2 \Omega}{\Sigma \frac{\partial \left(  r^2 \Omega \right)}{\partial r}} .
\end{equation}

\section{ disk Power Loss }
\label{sec:disk_Power_Loss}

As in \ref{sec:magnetic_braking}, we consider the magnetic braking action on an annular volume element. The total power, $\Delta P_D$, produced from the ring element is
\begin{equation}
\begin{aligned}
    &  \Delta P_D = \mathbf{F}_D \cdot \mathbf{v}_\theta \ \Delta V \\
    & = - 4 \pi \sigma r^3 \Omega^2 B^2 h \ \Delta r .
\label{eq:disk_element_power}
\end{aligned}
\end{equation}

In the limit of $\Delta r \to 0$

\begin{equation}
    \frac{\text{d}P_D}{\text{d}r} = - 4 \pi \sigma r^3 \Omega^2 B^2 h \ .
\label{eq:disk_differential_power}
\end{equation}
If we assume disk averaged values $\bar{\sigma}, \bar{h}$ and $\bar{B}$ plus Keplerian angular disk speed, ($\Omega \approx \sqrt{{\rm G}M_*/r^3}$), then the total power produced from the disk, $P_{D}$, is:
\begin{equation}
    P_{D} \approx - 4 \pi \bar{\sigma} \bar{B}^2 \bar{h} {\rm G} M_* (r_{max} - r_{min}) \ ,
\label{eq:disk_total_power}
\end{equation}
with $r_{max}$ and $r_{min}$, respectively, the maximum and minimum disk radii.

\section{Analytic Solution of the disk Equation }
\label{sec:analytic solution}

Returning to the mass conservation equation for a thin disk:

\begin{equation}
   r \frac{\partial \Sigma}{\partial t} = -\frac{\partial \left(r v_{\rm r} \Sigma \right)}{\partial r} \ .
    \label{eq:mass_con1}
\end{equation}

We can substitute in this equation for the radial infall speed:
\begin{equation}
    v_{\rm r} = -\frac{4 h\sigma r B^2}{\Sigma} \ .
\label{eq:rad_drift3}
\end{equation}

$\implies$
\begin{equation}
   \frac{\partial \Sigma}{\partial t} = \frac{1}{r}\frac{\partial \left(4h \sigma r^2 B^2 \right)}{\partial r} \ .
    \label{eq:mass_con2}
\end{equation}

Suppose, as an illustrative first example, we take representative or "disk-averaged" values: $\bar{\sigma}, \bar{h}$ and $\bar{B}$ as replacements for $\sigma, h,$ and $B$. These representative values are assumed to be constant in $r$ and $t$ and so:
\begin{equation}
   \frac{\partial \Sigma}{\partial t} \approx 8 \bar{h} \bar{\sigma} \bar{B}^2 \equiv \dot{\Sigma}_a \ ,
    \label{eq:mass_con3}
\end{equation}
where $\dot{\Sigma}_a$ is the rate of accretion surface mass density. For this illustrative example, we assume that $\dot{\Sigma}_a$ is constant in $r$ and $t$, but this would not be true in general.

Integrating the above equation, we obtain
\begin{equation}
   \Sigma(r,t) \approx \dot{\Sigma}_a (t-t_0) + \Sigma_0(r_0,t_0) \ ,
    \label{eq:disk_surface_density_evolution}
\end{equation}
with $\Sigma_0(r_0,t_0)$ the initial value of $\Sigma$ at the initial position and time, $r_0$ and $t_0$.

\section{ Radial Position as a Function of Time }
\label{sec:analytic radial solution}

Combining equations (\ref{eq:rad_drift3}) and (\ref{eq:disk_surface_density_evolution})

\begin{equation}
    \begin{aligned}
    v_{\rm r} = \frac{{\rm d} r}{{\rm d} t} & \approx -\frac{4 \bar{h}\bar{\sigma} r \bar{B}^2}{\dot{\Sigma}_a (t-t_0) + \Sigma_0} \\
    & -\frac{\dot{\Sigma}_a r}{2\left( \dot{\Sigma}_a (t-t_0) + \Sigma_0 \right)} \ .
    \end{aligned}
\label{eq:rad_drift4}
\end{equation}

$\implies$
\begin{equation}
     \frac{1}{r}\frac{{\rm d} r}{{\rm d} t}  \approx 
    -\frac{\dot{\Sigma}_a }{2\left( \dot{\Sigma}_a (t-t_0) + \Sigma_0 \right)} \ ,
\label{eq:rad_drift5}
\end{equation}

which has the solution

\begin{equation}
    \begin{aligned}
    r & \approx r_0 \left( \frac{ \Sigma_0}{ \dot{\Sigma}_a (t-t_0) + \Sigma_0 } \right)^{1/2} \\
      & =  \frac{r_0}{ \sqrt{1 + (t-t_0)/\tau}} \ ,
    \end{aligned}
\label{eq:radial position vs time}
\end{equation}

with

\begin{equation}
     \tau  = \frac{\Sigma_0}{\dot{\Sigma}_a} \ .
\label{eq:drift timescale}
\end{equation}

\end{appendix}

\end{document}